\documentclass[a4paper,11pt]{article}
\usepackage{pos}

\title{Recent Advances in Hadron Production in $e^+e^-$ Annihilation at BESIII}

\author*[a,b]{Xiongfei Wang}

\onbehalf{(on behalf of BESIII Collaboration)}

\affiliation[a]{School of Physical Science and Technology,
Lanzhou University,
Lanzhou 730000, People’s Republic of China
}
\affiliation[b]{Lanzhou Center for Theoretical Physics,
Frontiers Science Center for Rare Isotopes,
Key Laboratory of Theoretical Physics of Gansu Province,
Key Laboratory for Quantum Theory
and Applications of MoE, 
Lanzhou University,
Lanzhou 730000, People’s Republic of China}
\emailAdd{wangxiongfei@lzu.edu.cn}

\abstract{
This talk presents recent experimental findings at BESIII experiment, including three distinct studies. 
Firstly, it reports the measurements of the Born cross-sections of 
$e^+e^-\to\Sigma^+\bar\Sigma^-$, $\Xi^-\bar\Xi^+$, $K^{0}_{S}K^{0}_{L}$ and $K^-\bar\Xi^+\Lambda/\Sigma^0+c.c.$ at center-of-mass energies between 3.5 and 4.9 GeV.  Secondly, it delves into the observation of significant flavor-SU(3) breaking in the kaon wave function at 12 GeV$^2$$<Q^2<$25 GeV$^2$, the discovery of $\psi(3770)\to K^{0}_{S}K^{0}_{L}$ and the evidence of $\psi(4160)\to K^-\bar\Xi^+\Lambda+c.c.$. 
Thirdly, it discusses the search for the production of deuterons in $e^+e^-$ annihilation at center-of-mass energies between 4.1 and 4.7 GeV in the reactions of 
$e^+e^-\to pp\pi^-\bar{d}+c.c.$ and $pp\bar{p}\bar{n}\pi^-+c.c.$. 
These results offer new perspectives on hadron production in the baryonic final states and the flavor dynamics within the kaon wave function, and the rare decay processes involving the $\psi(3770)$ resonance, contributing to our understanding of hadron dynamics in this energy regime through baryon final states produced in $e ^+ e ^-$ collisions.
}

\FullConference{%
  *** The 10th International Conference on Quarks and Nuclear Physics (QNP2024), ***\\
  *** 8–12 Jul 2024 ***\\
  *** Institute of Cosmos Sciences of the University of Barcelona, Barcelona, Spain ***
}

\begin{document}
\maketitle

\section{Introduction}
Study of hadron production in $e^+e^-$ annihilations above open-charm threshold  is crucial for understanding the nature of charmonium-like states and testing non-perturbation theory of QCD. 
The overabundance of the vector-charmonium(-like) states and the discrepancies between the potential model prediction and experimental measurements 
provide a significant opportunity to probe exotic configurations of quarks and gluons. 
The discovery of vector-charmonium(-like) states, known as XYZ particles such as $X(3872)$, $Y(4260)$, $Z_{c}(3900)$, etc., in $e^+e^-$ annihilations into charmonium and light hadrons highlights the importance of studying baryonic final states where information is still limited.

\section{The BEPCII/BESIII experiment}
\begin{figure}[!htbp]
\begin{center}
\includegraphics[width=0.98\textwidth]{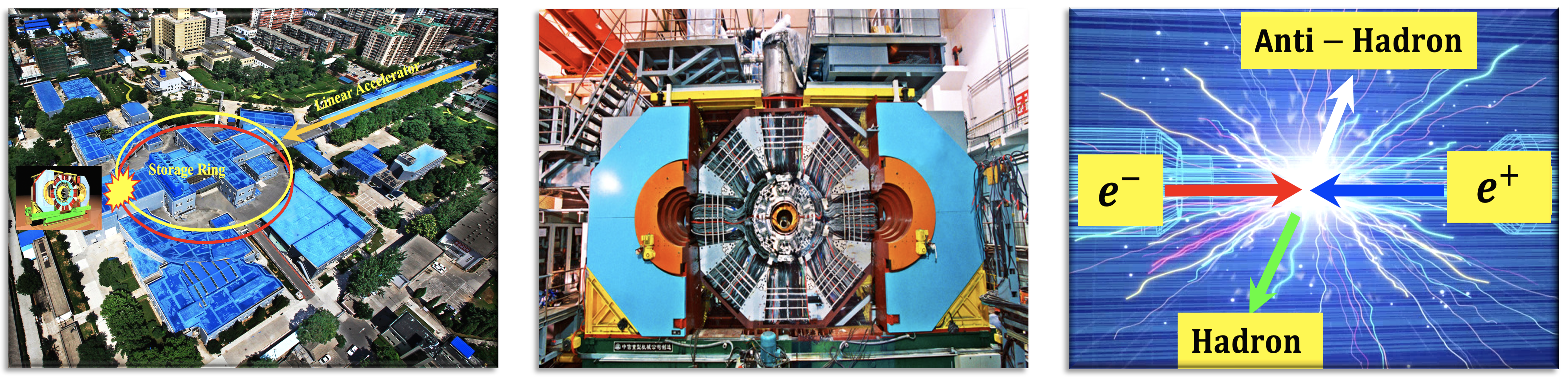}
\end{center}
\caption{\label{BEPC&&BESIII}Overview of BEPCIII (Left), BESIII detector (Middle) and the hadron pair production in $e^+e^-$ annihilation (Right).}
\end{figure}

The BEPCII is a double-ring $e^+e^-$ collider operating in the $\tau$-charm physics region with a design luminosity of $1\times 10^{33}\rm cm^{-2}s^{-1}$ at the beam energy of 1.89 GeV~\cite{Yu:2016cof}. It comprises a linear accelerator with a length of 200 m, a dual storage rings with a circumference of 240 m, and an BESIII detector situated at
in the southern end of the BEPCII as shown in Fig.~\ref{BEPC&&BESIII}. 
On April 5, 2016, BEPCII collider successfully achieved its design luminosity~\cite{Li:2021zrf}.
The BESIII detector~\cite{BESIII:2009fln} started data taking in 2009 and so far has accumulated huge data samples  in center-of-mass (CM) energies ($\sqrt{s}$) between 2 and 4.9 GeV~\cite{Ablikim:2013ntc,BESIII:2015qfd, BESIII:2017lkp,BESIII:2022ulv,BESIII:2024lbn}. 
\section{Recent advances}

\subsection{$e^+e^-\to\Xi^-\bar\Xi^+$}
In 2020, BESIII experiment performed a first search for the charmonium(-like) states, $Y(4230)$, $Y(4260)$ through an energy scan for the $e^+e^-\to\Xi^-\bar\Xi^+$ reaction in the CM energies between 4.0 and 4.6 GeV~\cite{BESIII:2019cuv}. No significant evidence for  $Y(4230)$ and $Y(4260)$ decaying into the $\Xi^-\bar\Xi^+$ final states was seen.
 Recently, BESIII experiment further improved the measurement of Born cross section and effective form factor for the $e^+e^-\to\Xi^-\bar\Xi^+$ reaction with more 
 thirty-two points in the CM energies between 3.5 and 4.9 GeV as shown in Fig.~\ref{Re:Ad:SigSig} (a).  
 A fit to the dressed cross section of the reaction was performed in combination with previous BESIII measurements~\cite{BESIII:2023rse}. Describing the energy-dependence cross section requires an evidence of   
 $\psi(3770)\to\Xi^-\bar\Xi^+$ with a significance of 4.5 $\sigma$, accounting for the systematic uncertainty as shown in Fig.~\ref{Re:Ad:SigSig} (b). For the other charmonium(-like) states $\psi(4040)$,  $\psi(4160)$,  $Y(4230)$, $Y(4360)$, $\psi(4415)$ or $Y(4660)$, no significant signal of their decay to the $\Xi^-\bar\Xi^+$ final states was found. Only upper limit for the product of the branching fraction and the electronic partial widths $\Gamma_{ee}{\cal{B}}$
at the 90\% C.L. was reported. This measurement provides the first evidence for charmless decays of 
$\psi(3770)$, and can be useful for the understanding of the charmonium(-like) states coupling to the baryon-antibaryon final states.
\subsection{$e^+e^-\to\Sigma^+\bar\Sigma^-$}
\begin{figure}[!htbp]
\begin{center}
\includegraphics[width=0.95\textwidth]{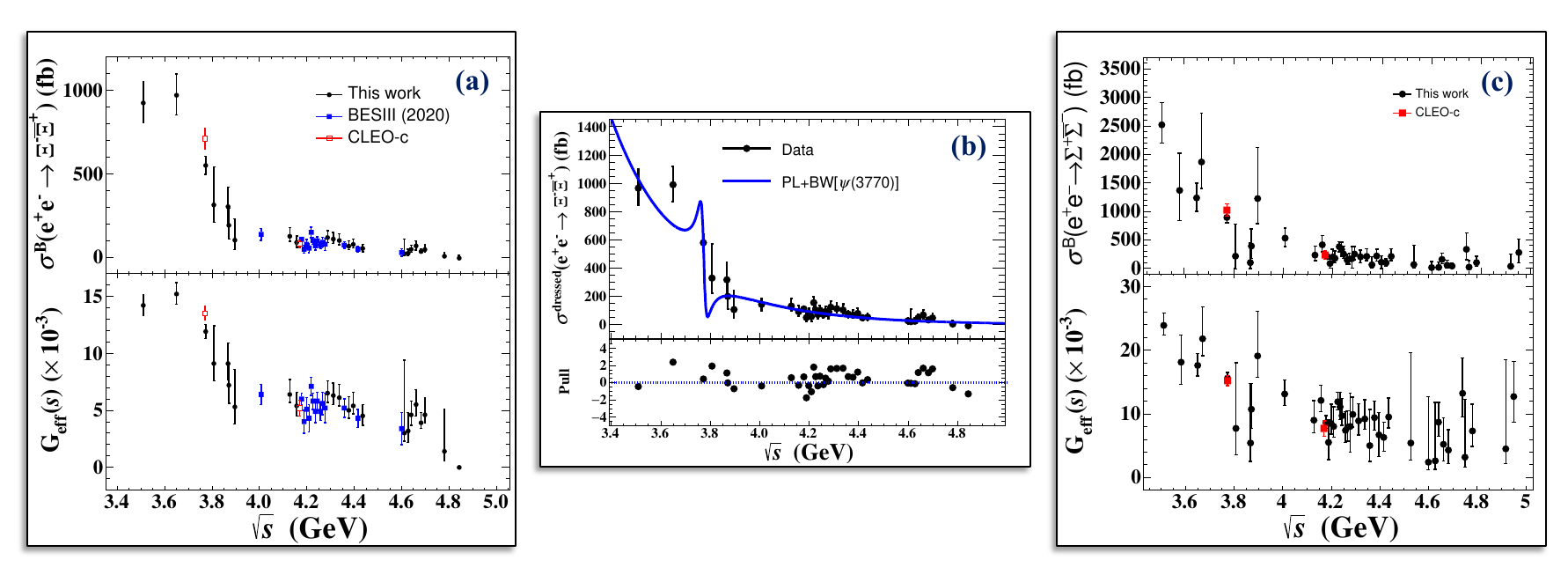}
\end{center}
\caption{\label{Re:Ad:SigSig} 
Comparisons of $\sigma^{\rm B}$ and $G_{\rm eff}(s)$ as a function of CM energy for $e^+e^-\to\Sigma^+\bar\Sigma^-$ between BESIII and CLEO-c experiment (Left) for $e^+e^-\to\Xi^-\bar\Xi^+$ between BESIII (2020), BESIII (2023) and CLEO-c experiment (Middle).
Fit to the dressed cross section with the different assumptions (Right).}
\end{figure}

Using a data sample corresponding an integrated luminosity of 24 fb$^{-1}$,  BESIII experiment conducted the first study of the $e^+e^-\to\Sigma^+\bar\Sigma^-$ reaction~\cite{BESIII:2024umc}.
The measured Born cross section and effective form factor with a total of forty-one points in CM energies between 3.5 and 4.9 GeV were reported, and compared with the results of CLEO-c experiment~\cite{Dobbs:2017hyd} at $\sqrt{s} =$ 3.770 and 4.160 GeV as shown in Fig.~\ref{Re:Ad:SigSig} (c). 
Potential charmonium(-like) states were also investigated by fitting the dressed cross section for the $e^+e^-\to\Sigma^+\bar\Sigma^-$ reaction with an assumption of a power-law (PL) function plus single resonance [i.e. $\psi(3770)$,  $\psi(4040)$,  $\psi(4160)$,  $Y(4230)$, $Y(4360)$, $\psi(4415)$ or $Y(4660)$] one at a time. No significant evidence for any assumed resonance was seen. Only the upper limits of $\Gamma_{ee}{\cal{B}}$ at the 90\% confidence level (C.L.) were evaluated. These results contribute more information for  understanding the nature of charmonium(-like) states above open charm threshold.   

\subsection{$e^+e^-\to K^{0}_{S}K^{0}_{L}$}
\begin{figure}[!htbp]
\begin{center}
\includegraphics[width=0.72\textwidth]{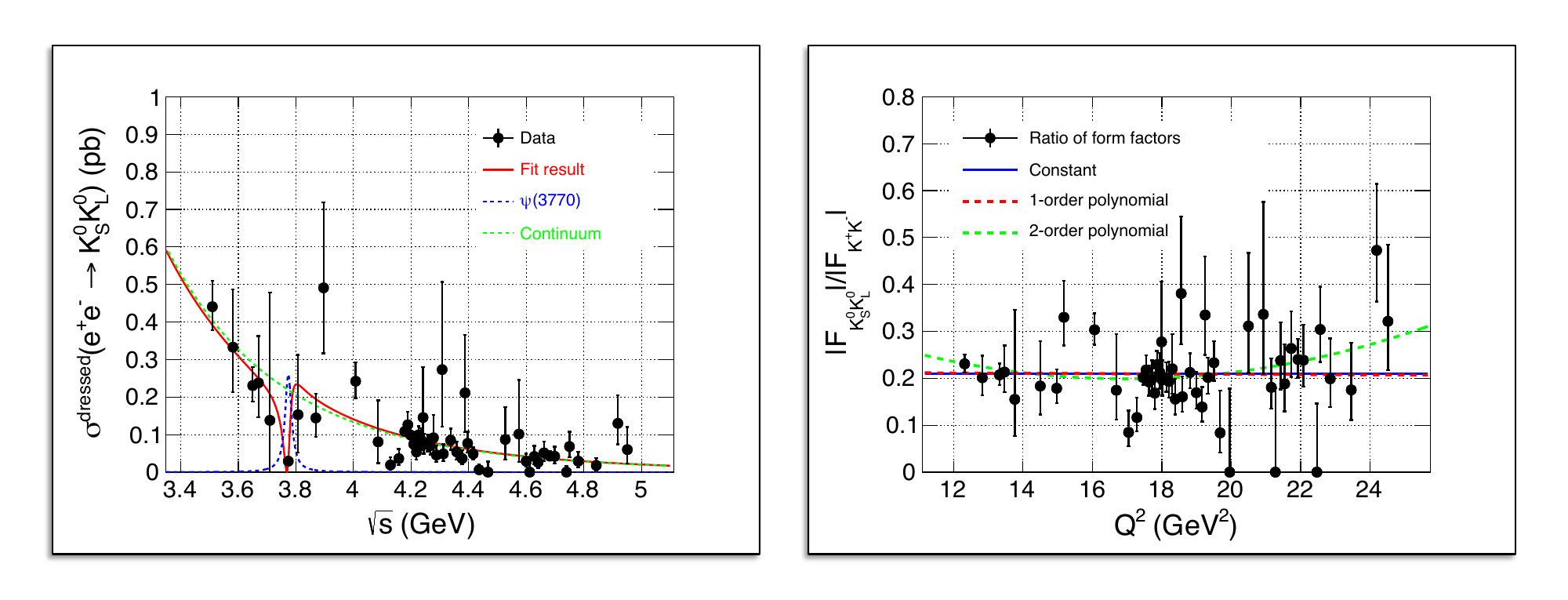}
\end{center}
\caption{\label{Re:Ad:KSKL} 
Fit to the dressed cross section for the $e^+e^-\to K^{0}_{S}K^{0}_{L}$ reaction (Left).
The ratio of neutral-to-charged kaon form factors (Right).}
\end{figure}

Using a data sample corresponding to an integrated luminosity of 27 fb$^{-1}$, BESIII experiment reported the first discovery of the charmless decay of the $\psi(3770)\to K^{0}_{S}K^{0}_{L}$ with a statistical significance of 10 $\sigma$ by an energy scan method~\cite{BESIII:2023zsk}.
Figure~\ref{Re:Ad:KSKL} shows the fit to the dressed cross section of $e^+e^-\to K^{0}_{S}K^{0}_{L}$ assuming a continuum contribution plus a $\psi(3770)$ resonances in the CM energy between 3.5 and 4.9 GeV.
The measured branching fraction for $\psi(3770)\to K^{0}_{S}K^{0}_{L}$ is ${\cal{B}} =
2.63^{+1.40}_{-1.59} \times 10^{-5}$,
which is consistent with the prediction of the S-and D-wave charmonia mixing model developed to interpret the $\rho\pi$ puzzle between $J/\psi$ and $\psi(3686)$ decays~\cite{Rosner:2001nm}.
  Besides, the ratio of neutral-to-charged kaon form factors at large momentum transfers in
 $12 < Q^2 < 15$ GeV$^2$, as shown in Fig.~\ref{Re:Ad:KSKL} was determined to be 
 $0.21 \pm 0.01$, indicating a small but significant effect of flavor-$SU(3)$ breaking in the kaon wave function. This result excludes the possibility 
 for the large deviation between the pQCD prediction~\cite{Lepage:1980fj} and CLEO-c result~\cite{Seth:2012nn}, and is inconsistent with  the predicted trend using a single bound-state interaction kernel~\cite{Gao:2017mmp}.
These findings provide valuable information for investigating the internal structure of neutral kaons.
\subsection{$e^+e^-\to K^-\bar\Xi^+\Lambda/\Sigma^0+c.c.$}
\begin{figure}[!htbp]
\begin{center}
\includegraphics[width=0.75\textwidth]{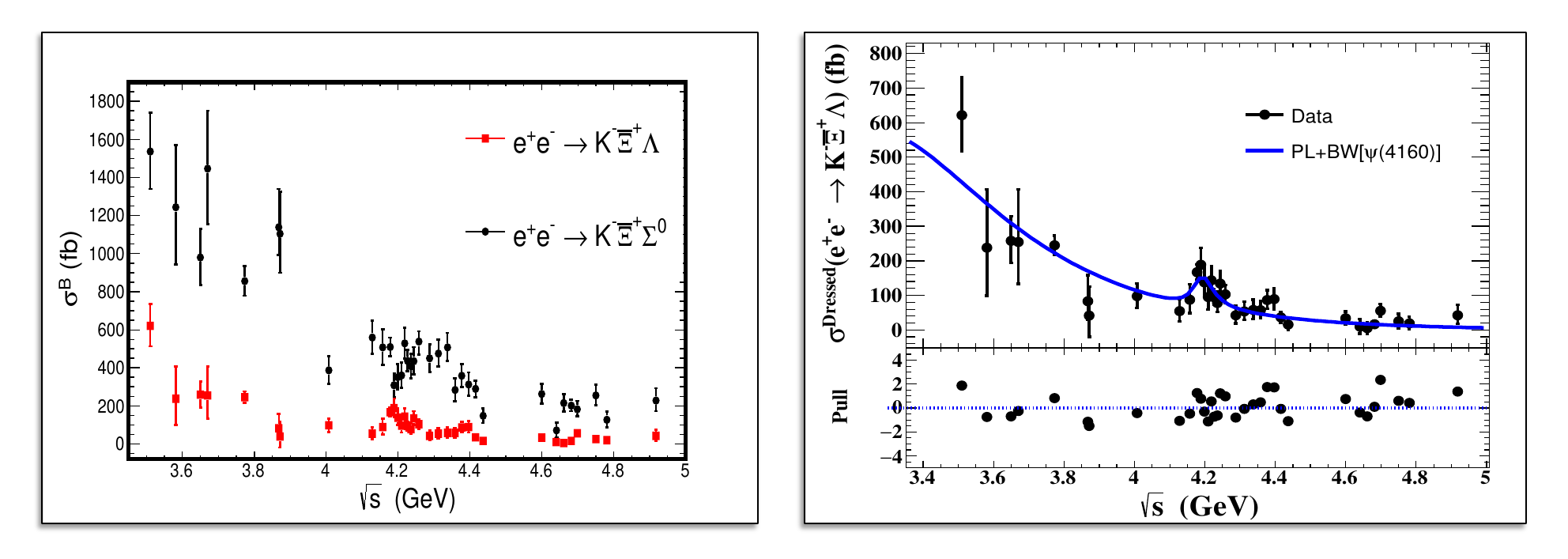}
\end{center}
\caption{\label{Re:Ad:KXiLam} 
The measured Born cross sections for the $e^+e^-\to K^-\bar{\Xi}^+\Lambda$ and $K^-\bar{\Xi}^+\Sigma^{0}$ reactions 
as a function of CM energy between 3.5 and 4.9 GeV (Left).
Fit to the dressed cross section with the assumption of $\psi(4160)$ resonance plus a continuum contribution.}
\end{figure}

BESIII experiment has also reported a measurement of the Born cross sections for the $e^+e^-\to K^-\bar{\Xi}^+\Lambda/\Sigma^{0}$ reaction at thirty-five CM energy points between 3.5 and 4.9 GeV~\cite{BESIII:2024ogz} as shown in Fig.~\ref{Re:Ad:KXiLam}.
Following  a fit to the dressed cross sections for  $e^+e^-\to K^-\bar{\Xi}^+\Lambda/\Sigma^{0}$ with the assumption of one resonance plus a continuum contribution, evidence for the decay of $\psi(4160)\to K^-\bar\Xi^+\Lambda$ with a significance of 4.4 $\sigma$ including systematic uncertainty is found as shown in Fig.~\ref{Re:Ad:KXiLam} (Right).
No significant evidence for any state decaying into the $K^-\bar\Xi^+\Lambda/\Sigma^0$ final state can be found.
The upper limits for $\Gamma_{ee}{\cal{B}}$ t the 90\% C.L.  for all assumed resonances decaying into the $K^-\bar{\Xi}^+\Lambda/\Sigma^0$ final state were determined. 
These results are valuable as they contribute to the  experimental information regarding the three-body baryonic decay of charmonium (-like) states.

\subsection{$e^+e^-\to pp\bar{d}\pi^-+c.c.$}
The investigation of hadronic states containing six quarks, such as compact hexaquark states or dibaryon states, remains a prominent area of research.
Among them, the $d(2380)^*$ has garnered significant attention. Recently, BESIII experiment conducted a search for the production of deuterons and
antideuterons through $e^+e^-\to pp\bar{d}\pi^-+c.c.$ at CM energies between 4.13 and 4.70 GeV~\cite{BESIII:2024pue}. 
No significant signal for potential deuteron or antideuteron candidates can be seen as shown in Figs.~\ref{Re:Ad:4ppi} (a) and (c). Only upper limit for the cross section of $e^+e^-\to pp\bar{d}\pi^-+c.c.$ is determined as shown in Fig.~\ref{Re:Ad:4ppi} (b).
\subsection{$e^+e^-\to pp\bar{p}\bar{n}\pi^-+c.c.$}
\begin{figure}[!htbp]
\begin{center}
\includegraphics[width=0.49\textwidth]{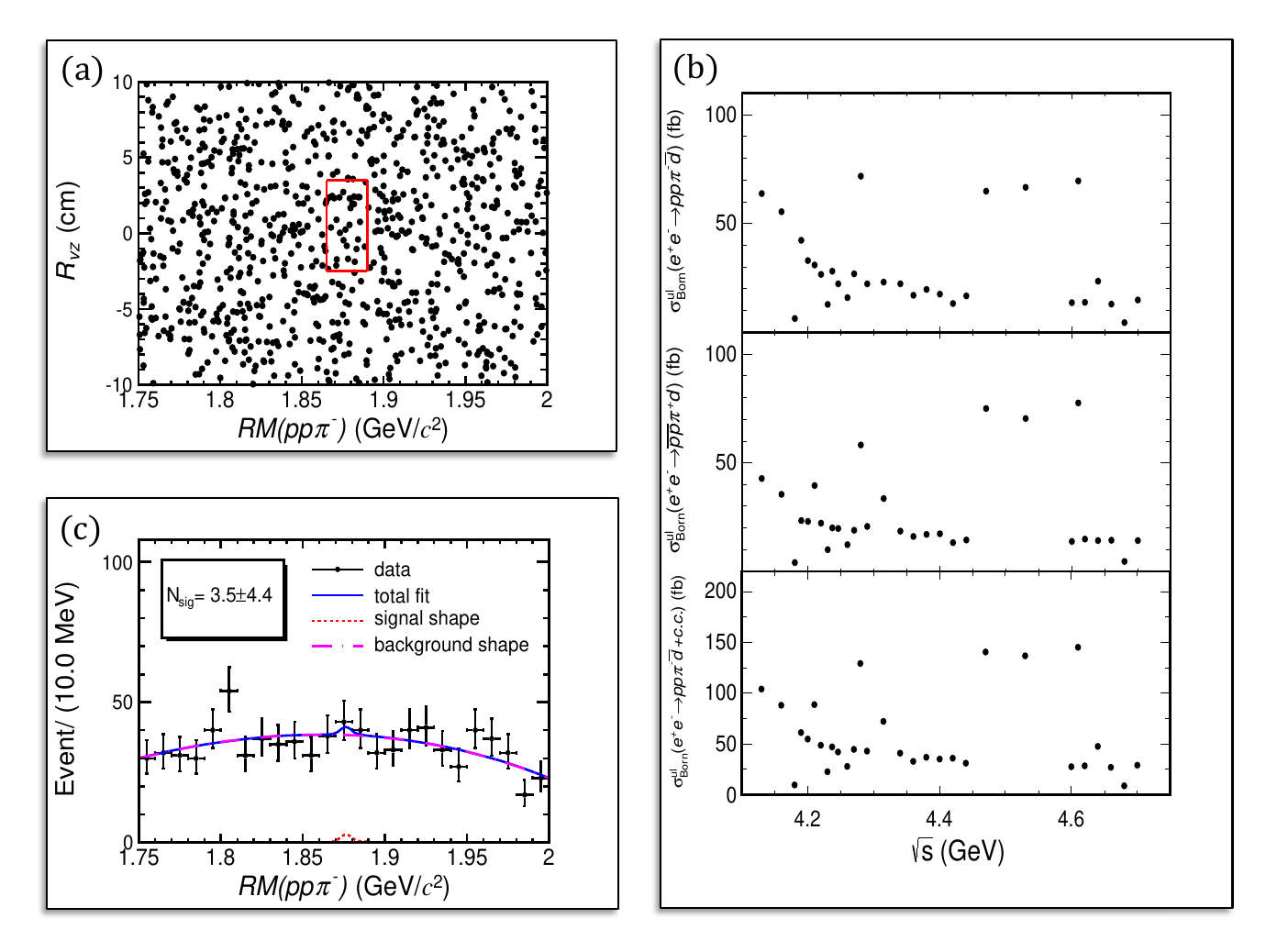}
\includegraphics[width=0.49\textwidth]{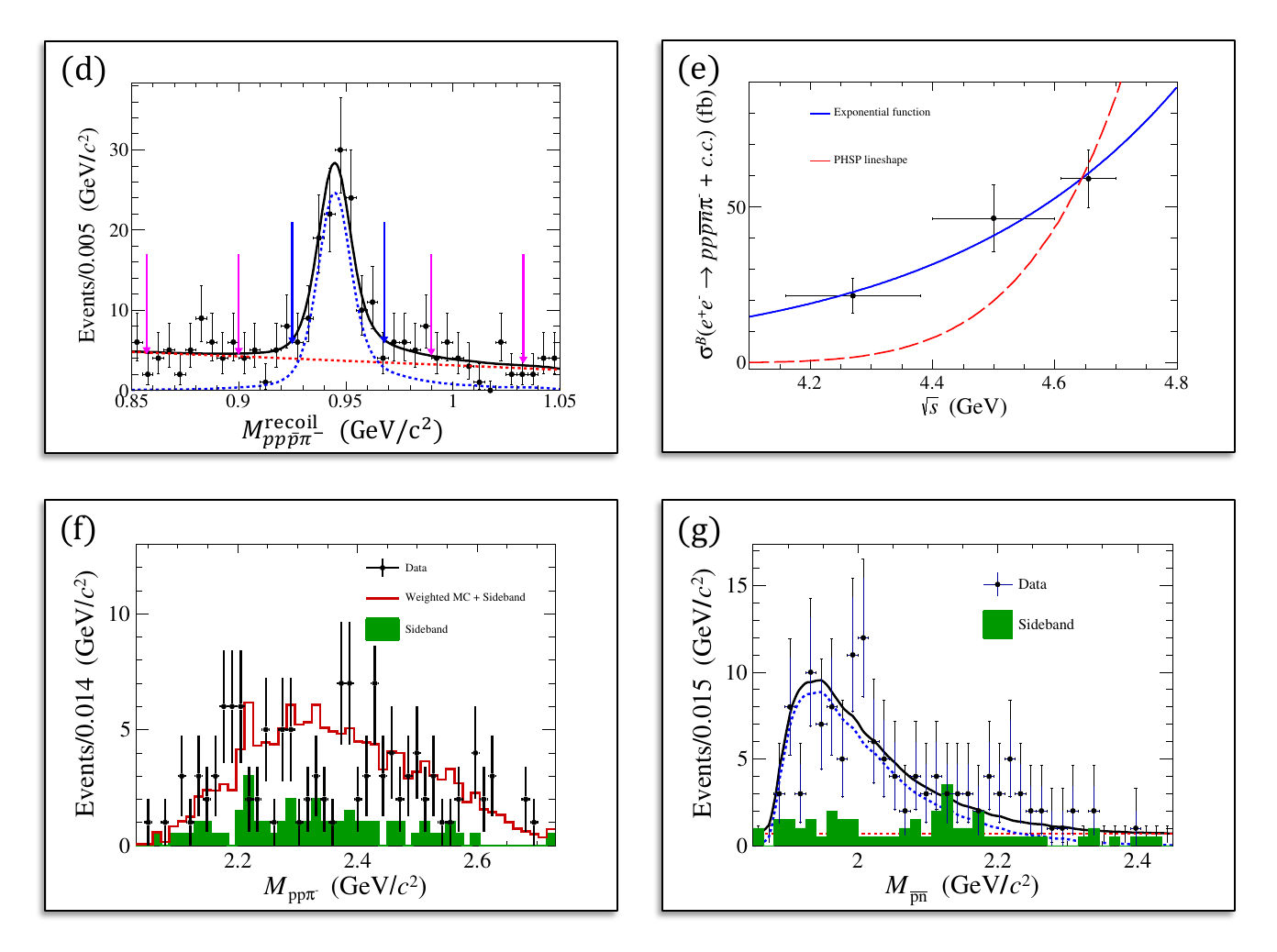}
\end{center}
\caption{\label{Re:Ad:4ppi} 
(a) Scatter plot of $R_{vz}$ versus $RM(pp\pi^-)$,  (b) upper limits of Born cross section and (c) distribution of $RM(pp\pi^-)$ for the $e^+e^-\to pp\bar{d}\pi^-+c.c.$ reaction.
(d) Distribution of $M^{\rm recoil}_{pp\bar{p}\pi^-}$, (e) the measured Born cross section at three subsamples, (f) distribution of $M_{pp\pi^-}$ and (g) distribution of $M_{p\bar{n}}$ for the $e^+e^-\to pp\bar{p}\bar{n}\pi^-+c.c.$ reaction.}
\end{figure}

No deuteron signal can be seen in the $e^+e^-\to pp\bar{d}\pi^-+c.c.$ reaction due to the large background. Consequently,  BESIII experiment subsequently performed a refined analysis by tracking additional final states via the  $e^+e^-\to pp\bar{p}\bar{n}\pi^-+c.c.$ reaction with only one missing antineutron~\cite{BESIII:2022acw}. Figure~\ref{Re:Ad:4ppi} (d) illustrates the recoil mass spectrum of $pp\bar{p}\pi^-$ by sum of all energy points. A clear neutron signal was observed with a significance of 11.5 $\sigma$. The average Born cross sections with three energy ranges were measured as shown in Fig.~\ref{Re:Ad:4ppi} (e), 
revealing discrepancies with the expectation from five-body phase space (PHSP).
The shapes of $M_{pp\pi^-}$ and $M_{p\bar{n}}$ align well the PHSP distributions as shown in Figs.~\ref{Re:Ad:4ppi} (f) and (g), which thereby indicates no hexaquark or di-baryon state can be observed under the current data sample size.

\section{Summary}
The BESIII experiment has been operating successfully since 2008 and has collected the largest data samples in the $\tau$-charm physics region. 
Many advances in hadron production in $e^+e^-$ annihilation at BESIII has been achieved, including the evidences of $\psi(3770)\to\Xi^-\bar\Xi^+$ and $\psi(4160)\to K^-\bar\Xi^+\Sigma^0 +c.c.$, observation of $\psi(3770)\to K^{0}_{S}K^{0}_{L}$, search for the deuteron and so on. 
These findings offer valuable insights into the production of the baryon final states, contributing to our understanding of hadron dynamics within this energy regime and shedding light on the mechanisms and properties of light nuclei, thus enhancing our comprehension of hadronization dynamics and quark-gluon interactions.

\section{Acknowledgement}
This work is supported in part by National Key Research and Development Program of China under Contracts Nos. 2020YFA0406400, 2020YFA0406403; National Natural Science Foundation of China under Contracts Nos. 12075107, 12247101; the 111 Project under Grant No. B20063.


\begin{thebibliography}{99}


\bibitem{Yu:2016cof}
C.~Yu \textit{et al.},
\href{https://inspirehep.net/files/96083dc6a03caf20597ac55b4500673a}{Proceedings of IPAC2016}, Busan, Korea, 2016.
\bibitem{Li:2021zrf}
G.~Li \textit{et al.},
\href{https://epjtechniquesandinstrumentation.springeropen.com/articles/10.1140/epjti/s40485-021-00064-9}{EPJ Tech. Instrum. \textbf{8}, 7 (2021)}.
\bibitem{BESIII:2009fln}
BESIII Collaboration,
\href{https://www.sciencedirect.com/science/article/pii/S0168900209023870?via\%3Dihub}{Nucl. Instrum. Meth. A \textbf{614}, 345-399 (2010)}.
\bibitem{Ablikim:2013ntc}
BESIII Collaboration,
\href{https://iopscience.iop.org/article/10.1088/1674-1137/39/9/093001}{Chin. Phys. C \textbf{37}, 123001 (2013)}.
\bibitem{BESIII:2015qfd}
BESIII Collaboration,
\href{https://iopscience.iop.org/article/10.1088/1674-1137/39/9/093001}{Chin. Phys. C \textbf{39}, 093001 (2015)}.
\bibitem{BESIII:2017lkp}
BESIII Collaboration,
\href{https://iopscience.iop.org/article/10.1088/1674-1137/41/6/063001}{Chin. Phys. C \textbf{41}, 063001 (2017)}.
\bibitem{BESIII:2022ulv}
BESIII Collaboration,
\href{https://iopscience.iop.org/article/10.1088/1674-1137/ac80b4}{Chin. Phys. C \textbf{46}, 113003 (2022)}.
\bibitem{BESIII:2024lbn}
BESIII Collaboration, \href{https://arxiv.org/abs/2406.05827}{[arXiv:2406.05827 [hep-ex]]}.



\bibitem{BESIII:2024umc}
BESIII Collaboration,
\href{https://link.springer.com/article/10.1007/JHEP05(2024)022}{JHEP \textbf{05}, 022 (2024)}.
\bibitem{Dobbs:2017hyd}
S.~Dobbs \textit{et al.},
\href{https://journals.aps.org/prd/abstract/10.1103/PhysRevD.96.092004}{Phys. Rev. D \textbf{96}, 092004 (2017)}.


\bibitem{BESIII:2019cuv}
BESIII Collaboration,
\href{https://journals.aps.org/prl/abstract/10.1103/PhysRevLett.124.032002}{Phys. Rev. Lett. \textbf{124}, 032002 (2020)}.


\bibitem{BESIII:2023rse}
BESIII Collaboration,
\href{https://link.springer.com/article/10.1007/JHEP11(2023)228}{JHEP \textbf{11}, 228 (2023)}.


\bibitem{BESIII:2023zsk}
BESIII Collaboration,
\href{https://journals.aps.org/prl/abstract/10.1103/PhysRevLett.132.131901}{Phys. Rev. Lett. \textbf{132}, 131901 (2024)}.
\bibitem{Rosner:2001nm}
J.~L.~Rosner,
\href{https://journals.aps.org/prd/abstract/10.1103/PhysRevD.64.094002}{Phys. Rev. D \textbf{64}, 094002 (2001)}.


\bibitem{Lepage:1980fj}
G.~P.~Lepage and S.~J.~Brodsky,
\href{https://journals.aps.org/prd/abstract/10.1103/PhysRevD.22.2157}{Phys. Rev. D \textbf{22}, 2157 (1980)}.
\bibitem{Seth:2012nn}
K.~K.~Seth \textit{et al.},
\href{https://journals.aps.org/prl/abstract/10.1103/PhysRevLett.110.022002}{Phys. Rev. Lett. \textbf{110}, 022002 (2013)}.
\bibitem{Gao:2017mmp}
F.~Gao \textit{et al.},
\href{https://journals.aps.org/prd/abstract/10.1103/PhysRevD.96.034024}{Phys. Rev. D \textbf{96}, 034024 (2017)}.


\bibitem{BESIII:2024ogz}
BESIII Collaboration,
\href{https://link.springer.com/article/10.1007/JHEP07(2024)258}{JHEP \textbf{07}, 258 (2024)}.

\bibitem{BESIII:2024pue}
BESIII Collaboration,
\href{https://journals.aps.org/prd/abstract/10.1103/PhysRevD.109.092013}{Phys. Rev. D \textbf{109}, 092013 (2024)}.

\bibitem{BESIII:2022acw}
BESIII Collaboration,
\href{https://iopscience.iop.org/article/10.1088/1674-1137/acb6eb}{Chin. Phys. C \textbf{47}, 043001 (2023)}.






\end{thebibliography}
\end{document}